# Water-filled telescopes[1]

Elio Antonello
*INAF-Osservatorio Astronomico di Brera*
elio.antonello@brera.inaf.it

**Abstract.**
In the present short note we discuss the case of the thought experiments on water-filled telescopes and of their realizations during 18th and 19th century.

## 1. Introduction

Several authors in recent years discussed in some detail the studies of the light propagation in water made in past centuries, and the different opinions on the effective importance of such tests. In the following we will refer mainly to those works.

Let us start by recalling the main ideas about light propagation in seventeenth century. The law of refraction was well established and accepted, but there were two different interpretations. According to the corpuscular or emission theory, the velocity of light was larger in the denser medium ($c_w$) than in air ($c$), i.e. $c_w > c$. Usually it is supposed this was the idea (or hypothesis) of Newton, though he declared at the beginning of his treatise that "My design in this book is not to explain the properties of light by hypotheses, but to propose them by reason and experiments" (Newton, 1704). However, he criticized some interpretations based on the wave theory (Pedersen, 2000). In the wave theory the velocity in the medium was smaller than in air, i.e. $c_w < c$, and it was believed by many scientists that a determination of the velocity of light in various media would have decided between the two theories.

## 2. Before the year 1800

*2.1. Melvill and Bradley*
After the celebrated discovery of stellar aberration by Bradley in 1729,

$$\tan \alpha = v/c,$$

where $\alpha$ is the aberration angle and $v$ the velocity of Earth, Melvill proposed in 1753 to test whether different colours are propagating with different velocities. This was easily tested by estimating the emersion times from eclipse of Jupiter satellites, and the verification showed indeed they were not different. Moreover he proposed to verify by accurate measurements that the velocity of light in empty space was actually different from the accepted one, since in Bradley's formula *c* was the light velocity in the eye aqueous humour. Of course, Bradley did not appreciate at all this proposal; it was nonsense, since in the stellar aberration the inclination of the tube has nothing to do with the light transmission in the eye of the observer (Pedersen, 2000).

---

[1] Paper presented at the International Conference for the tercentenary of the birth of Ruggiero Giuseppe Boscovich (Ragusa 1711 - Milano 1787), University of Pavia, Pavia, Italy September 8th - 10th, 2011.

*2.2 Boscovich*
Boscovich made a proposal to verify the emission theory by means of the measurement of the stellar aberration angle with two telescopes, one with air and the other with water. He expected a smaller aberration in water, since according to the emission theory the velocity was higher in water. He discussed this point in a letter to G.B. Beccaria in 1766; the letter however has not been found. He sent also a letter to the French Academy in 1770 with the project of the water telescope (correspondence with Lalande). Moreover, his *Risposta* to Kaunitz of 1772 included the proposal of an expensive quadrant with two telescopes, one of them water-filled, and he recommended this instrument for the Brera Observatory. The project was not approved. Lalande then published Boscovich's ideas in the fourth volume of his *Traité d'Astronomie* in 1781. Boscovich thought that by measuring the different aberration it was possible to determine the different velocity of light (Proverbio, 1993; Hollis, 1937).

*2.3 Wilson, Boscovich and Robison*
In England, Wilson proposed in 1770 to make a more accurate determination of *c* in empty space, as already proposed by Melvill. Then, in 1772, thinking about a water-filled telescope rather than the eye, he became convinced that his first reasoning was wrong, and a water (or any fluid) telescope would show the same stellar aberration as a telescope filled with air, and if the agreement between water and air telescope were found, it would have been a proof of the acceleration of the light in the dense medium, in the ratio assigned by the emission theory. Wilson published his study in 1782 (Pedersen, 2000). Up to this time, Boscovich was not aware of Wilson's proposal, nor Wilson was aware of Boscovich's project.

Boscovich in 1783 prepared the *Opera pertinentia* with an opusculum dedicated to the topic dealt with in the previous letter to Beccaria. He included also a new proposal for the measurement of terrestrial aberration with a water telescope, by observing a distant object. Some attempts to measure the terrestrial aberration were then performed by De Cesaris in the years 1784-1786 in the corridors of Brera Observatory with a water-filled telescope, according to the instructions that Boscovich sent him by letter. They gave unsatisfactory results due to technical problems related to the bad quality of the materials (Proverbio, 1993). During this period (1785), De Cesaris informed Boscovich about Wilson paper, but probably Boscovich never became acquainted with the details of such a work. It seems that his main worry was to defend his priority about water-filled telescopes both for stellar and terrestrial aberration measurements. After the publication of the *Opera* in 1785,

Wilson understood what was wrong in Boscovich's theory of terrestrial aberration, since it was the same thing that went wrong for himself in 1770 (Figure 1).

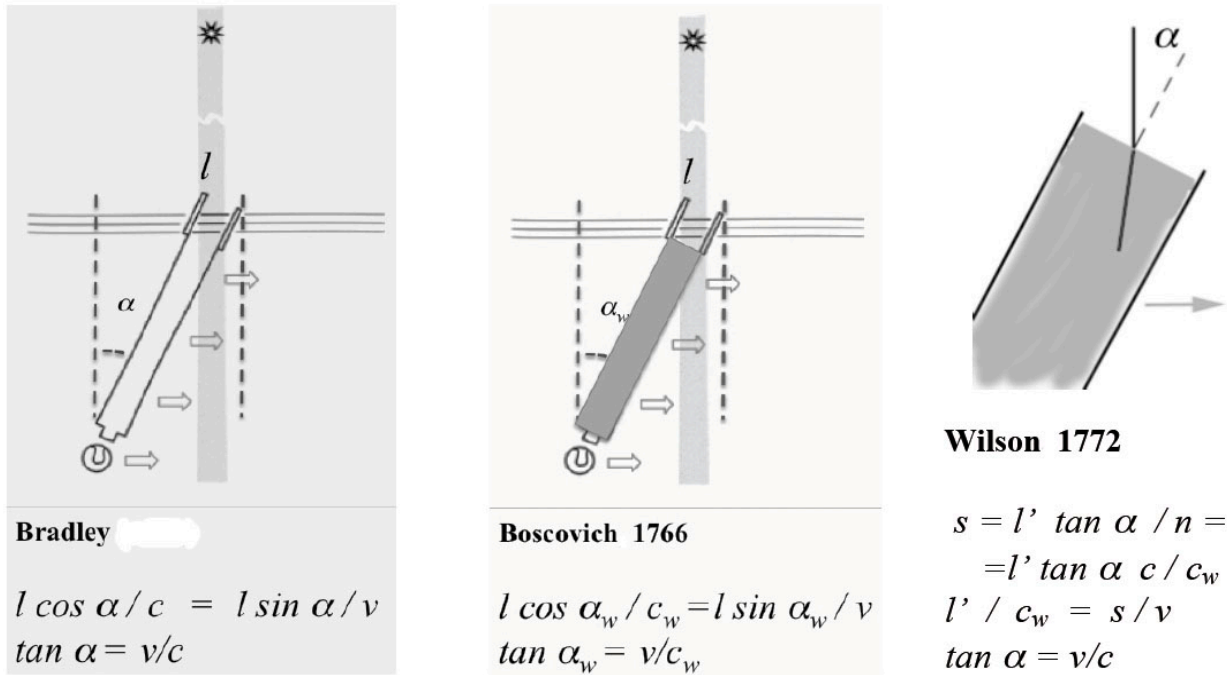

**Fig. 1.** *Left panel* (adapted from Norton, 2007): the telescope is moving with the earth; $l$ is the length of the tube. In the same time interval, while the light goes from the top to the bottom of the telescope, the telescope moves from the left to the same point. This is the geometry at the basis of Bradley's formula. *Central panel*: according to Boscovich, when the water is added, the geometry is the same, but the velocity is different ($c_w > c$), and the aberration angle is smaller. *Right panel*: according to Wilson, one has to take into account also the refraction of light in water; the ratio of the sinus of incidence and refraction angle is the refractive index $n$, $c_w > c$ and $c_w = c\,n$ in the corpuscular hypothesis; $l'$ is the path of the light in the tube, and $s$ is the displacement of the telescope (see Pedersen, 2000). In this case the conclusion is that the aberration with the water is the same as that without it.

Robison in the years 1785-1790 made some unsuccessful attempts to build a telescope filled with a sufficiently transparent substance for stellar aberration measurements, and then he devised a simpler instrument for measuring the terrestrial aberration. However, Robison realized that neither such an instrument nor the water-filled telescope would produce results different from ordinary instruments and telescopes. Therefore Robison never constructed his simple instrument (Pedersen, 2000).

*2.4 Conclusion*
The following conclusion seems evident: if the experiment devised by Boscovich had been performed reliably and accurately during the 18th century, the result would have been interpreted, after some discussion, as a confirmation of the emission

(corpuscular) theory of light, according to the reasoning of Wilson and Robison, that was based on the theory of refraction and the galilean relativity.

## 3. Between 1800 and 1860

### *3.1. Arago and Fresnel*
Arago expected to detect variations in the velocity of light from different stars owing to their different velocity, according to the emission theory. However, he did not found variations, and he concluded that any star emits light particle with any velocity, but only those particle with a velocity in a narrow interval can be detected by the human eye (Pedersen, 2000). Fresnel was interested in the wave theory of light and in the stellar aberration; however, Arago advised him to study the diffraction of light, and Fresnel wrote his important memory on the diffraction in 1818. Then Fresnel went back to the aberration, and he introduced the ether dragging theory. By the wave theory of light, a water-filled telescope would exhibit the same aberration of an ordinary telescope; physically, according to Fresnel, only part of the ether inside the moving body participate at the motion (dragging effect; Whittaker, 1910; Pedersen, 2000). In 1851 Fizeau made the experiment to measure the relative speeds of light in moving water, and he confirmed Fresnel's drag coefficient.

### *3.2. Conclusion*
By the mid of 19th century the thought experiments with water-filled telescopes to verify stellar and terrestrial aberration showed that the practical realizations of such experiments were not strictly required because they were not able to discriminate between emission theory and wave theory. However not everybody was convinced, and few people, as we will see, had a complete knowledge of the literature on this subject in order to be convinced. Moreover, understanding the nature of light required more experiments, since the more important an hypothesis, the larger the number of experiments to be performed in order to leave no doubt.

## 4. After 1860

### *4.1. Respighi, Klinkerfues and Hoek*
Respighi in Bologna considered Boscovich's ideas about terrestrial aberration, and he devised an experiment to measure reliably it, with the purpose to prove that Boscovich's expectations based on the emission theory were not correct, and to confirm in this way the wave theory of light (Gualandi & Bonoli, 2003). Apparently, he was not aware of the studies of Wilson and Robison. His accurate experiment (in 1859-1861) for the measurement of the terrestrial aberration did not show any effect due to the earth rotation on the observed target, and therefore he supported Fresnel's hypothesis (Whittaker, 1910).

Klinkerfues (1867) made a study of the refrangibility of light as affected by the motion of the source, of the possible dependence of the aberration constant on the passage of light through a refractory medium placed in the telescope and on the thickness of the objective glass. There had been a debate on the reason for the difference of the aberration constant measured by Delambre (20".255) and by Struve (20".445; the present day value is 20".495). Klinkerfues ascribed the difference to the objective used by Struve. A verification was made of the stellar aberration with a water-filled telescope in comparison with an ordinary one: according to Klinkerfues the stellar aberration in the water-filled telescope was increased by 7". Apparently, Klinkerfues, as Respighi, took into account the studies of Boscovich, but he was not aware of Wilson and Robison. Given the constraints on the use of the available instrumentation, Klinkerfues could make just few measurements and checks.

Hoek (1867) criticized the theory behind the experiments. He suggested that Klinkerfues did not apply correctly the Fresnel drag coefficient in the transversal component of the motion of the light in the telescope. Respighi considered the experiment of Klinkerfues inconclusive (Gualandi & Bonoli, 2003).

*4.2 Airy*

Airy (1871) was aware both of Boscovich and Wilson, therefore he was struck by the unexpected result of Klinkerfues. "A result of physical character so important, and resting on the respectable authority of Professor Klingerfues, merited and indeed required further examination". Therefore a verification was needed, and its result would have been "of great physical significant importance, not only affecting the computation of the velocity of light, but also influencing the whole treatment of the Undulatory Theory of Light." This is a strong statement by Airy, but it is justified only by the odd results of Klinkerfues. As it is well known, Airy found no difference in the stellar aberrations.

When Schiaparelli (1938) wrote a history of the contribution of Boscovich to the Brera Observatory, he included a curious comment: a water-filled telescope more than a century ago "would have been of major importance", and would have given matter of deep thoughts to the supporters of emission theory. As we have seen, this is not plausible. Apparently, Schiaparelli was not aware of Wilson and Robison, and probably his reasoning was based just on the strong sentence of Airy, which has a sense only if referred to Klinkerfues' results.

*4.3 Conclusion*

Apparently, complete information about past studies would have not justified the experiments done by Respighi and by Klinkerfues, and the consequent experiment by Airy. The experimental results should have been considered at most a confirmation of Fresnel's predictions (Whittaker, 1910). However, Airy experiment is usually deemed very important and it is quoted in the books of special relativity.

**5. Special Relativity**

Pauli (1958) commented Airy's result in his book. The earlier theory (Lorentz) had to make use of rather involved arguments in order to explain it, because it had to describe the effect as seen from a reference system relative to which the observer is moving. If it is observed from the rest system, it is self-evident from the relativistic point of view. "For if the telescope is pointed towards the apparent position of the fixed star, then the light waves sent out by it will have normal incidence on the telescope. If it is now filled with water, the light waves will be propagated normal to the boundary surface also in water. The Airy experiment, as seen from the rest system of the observer (earth), therefore only demonstrates the (relativistically) trivial fact that for a zero angle of incidence (normal incidence) the angle of refraction is zero, too" (Pauli, 1958).

According to Norton (2007), quoting Shankland (1963, 1973): "Prof. Einstein volunteered a rather strong statement that he had been more influenced by the Fizeau experiment … and by astronomical aberration especially Airy's observation with a water filled telescope, than by the Michelson-Morley experiment". "… the experimental results which had influenced him most were the observations of stellar aberration and Fizeau's measurements…"

The different opinion about the Airy experiment expressed by Pauli and by Einstein may be explained by the fact that, in the context of the special relativity as a well established theory, the result should be considered obvious, while it was no so when the theory was in the making.

## 6. Conclusion

The story of the water-filled telescope shows that the scientific progress occurs in a curious way. There was no stringent reason and no theoretical motivation for the construction of a water-filled telescope, but the experiment had been conceptually important for Einstein.